\documentclass[a4paper]{PoS}
\pdfoutput=1
\usepackage{amsmath}
\usepackage{slashed}

\newcommand{\ve}{\mathbf}
\newcommand{\U}{\text{U}}
\newcommand{\SU}{\text{SU}}

\newcommand{\refc}[1]{(\ref{#1})}

\title{QCD spectroscopy and quark mass renormalisation in external magnetic
fields with Wilson fermions\thanks{{\bf Acknowledgments:} The simulations have
been done on `iDataCool' at the university of Regensburg. This work was
supported by SFB/TRR-55. We would like to thank Kalman Szabo and Falk
Bruckmann for useful discussions.}}

\ShortTitle{Spectroscopy and quark mass renormalisation in ext. magn.
fields with Wilson fermions}

\author{Gunnar Bali, \speaker{Bastian B. Brandt},
   Gergely Endr\H{o}di and Benjamin Gl\"a\ss{}le \\
  Institute for Theoretical Physics\\
  University of Regensburg \\ D-93040 Regensburg \\
  E-mail: \email{bastian.brandt@ur.de}}

\abstract{We study the change of the QCD spectrum of low-lying mesons in the
presence of an external magnetic field using Wilson fermions in the quenched
approximation. Motivated by qualitative differences observed in the spectra of
overlap and Wilson fermions for large magnetic fields, we investigate the
dependence of the additive quark mass renormalisation on the magnetic field. We
provide evidence that the magnetic field changes the critical quark mass both in
the free case and on our quenched ensemble. The associated change of the bare
quark mass with the magnetic field affects the spectrum and is relevant for the
magnetic field dependence of a number of related quantities. We derive Ward 
identities for lattice and continuum QCD+QED from which we can extract the
current quark masses. We also report on a first test of the tuning of the quark
masses with the magnetic field using the current quark masses, and show that
this tuning resolves the qualitative discrepancy between the Wilson and overlap
spectra.}

\FullConference{The 33rd International Symposium on Lattice Field Theory\\
         14-18 July  2015\\
         Kobe International Conference Center, Kobe, Japan}

\begin{document}

\section{Introduction}

Besides the strong interest in studying the impact of QED on QCD observables
(for a review see e.g.~\cite{Portelli:2015wna}), there is also a growing
interest in the properties of QCD in strong external magnetic fields. They may
appear in non-central heavy-ion collisions~\cite{Kharzeev:2004ey}, inside
magnetars~\cite{Ferrer:2012wa} and in the evolution of the early
universe~\cite{Vachaspati:1991nm}. It is important to note that external
magnetic fields change the thermodynamic properties of QCD, as well as the
spectrum and other zero temperature characteristics. In particular, the change
in the energy levels of low-lying hadrons has an impact on all the physical
situations mentioned above. Furthermore, it was suggested~\cite{Schramm:1991ex}
that vector mesons might become massless and condense at some critical magnetic
field, leading to a superconductivity of the QCD vacuum along the magnetic field
axis~\cite{Chernodub:2011mc}. While the thermodynamic properties of QCD
in the presence of external magnetic fields are by now rather well understood
(see~\cite{Szabo:2014iqa}) there are only a few initial quenched studies of the
spectrum~\cite{Hidaka:2012mz,Luschevskaya:2014lga,Beane:2014ora,Detmold:2015daa}
. These studies suggest that $\rho$-meson condensation does not occur, although
indications for the contrary have also been reported~\cite{Braguta:2011hq}.

\begin{figure}[t]
 \centering
 \vspace*{-4mm}
\includegraphics[]{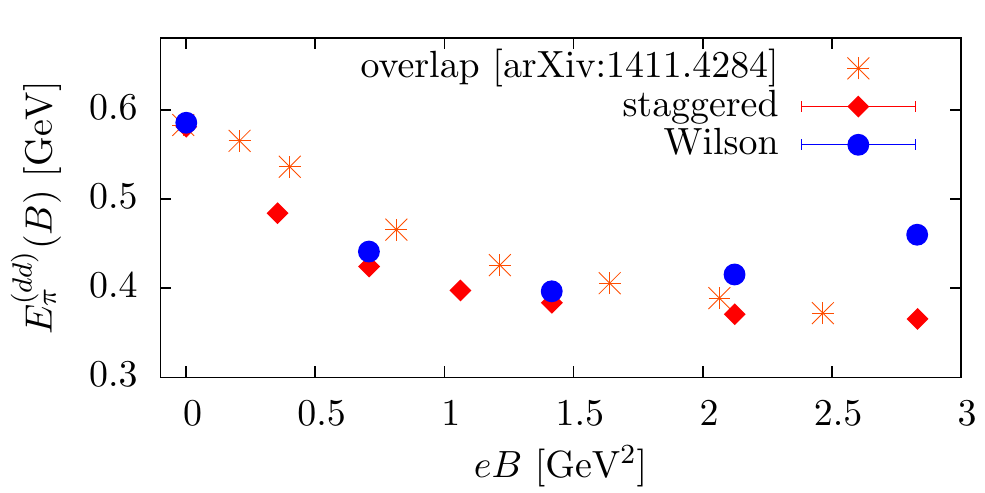}
 \vspace*{-1mm}
 \caption{Comparison between results for the energies of connected neutral pions
with $\bar{d}d$ flavour content with respect to the external magnetic field
obtained from different fermion discretisations. The results for Wilson and
staggered fermions are from the current study and the results for overlap
fermions are the taken from \cite{Luschevskaya:2014lga}. Here the pion mass at
$\ve{B}=\ve{0}$ is about 580 MeV.}
 \label{fig:pion-comp}
\end{figure}

In our study we aim at investigating the spectrum of low-lying hadrons exposed
to external magnetic fields $\ve{B}$ using Wilson fermions, initially in the
quenched setup and neglecting $O(a)$ improvement. Figure~\ref{fig:pion-comp}
shows a comparison of the results for the energies of the connected neutral
pions with $\bar{d}d$ flavour content, $\pi_{dd}$, obtained from Wilson and
staggered fermions on our quenched ensemble and overlap fermions
from~\cite{Luschevskaya:2014lga} at a $\ve{B}=\ve{0}$ pion mass of about
580~MeV and constant input quark mass $a\bar{m}$ (or constant hopping parameter
$\kappa=1/(2a\bar{m}+8)$). In contrast to the results from other fermion
discretisations, the results for Wilson fermions show a non-monotonous behaviour
with increasing $\ve{B}$. A crucial difference between Wilson fermions and the
other discretisations is the presence of an additive quark mass renormalisation
$a\bar{m}_{c}$. Its value depends on the properties of the Wilson term in the
Dirac operator where the magnetic field enters directly via the link variables,
see eq.~\refc{eq:u3-links}. Consequentially, it is to be expected that
$a\bar{m}_{c}$ will change with $B$, in analogy to its change in the presence of
QED~\cite{Portelli:2015wna}. In fact, this change of $a\bar{m}_{c}$, and
consequently the change in the bare quark masses $am_f\sim
a(\bar{m}_f-\bar{m}_{c,f})$, is already present in the free case, which we will
show in section~\ref{sec:free}. Though formally a lattice artefact, this effect
may be numerically large and, affecting the neutral pion energies additively,
its relative impact is enhanced towards small pion masses. This explains why the
effect is comparably small in figure~\ref{fig:pion-comp}, while it is much more
prominent below in figure~\ref{fig:qmren} (right) where the quark mass is
smaller. Moreover, as this effect results in quark masses to change with the
magnetic field at fixed $a\bar{m}$, quark mass and magnetic field dependence
become coupled in a non-trivial way. The impact of this subtlety of Wilson
fermion formulations on results obtained at finite lattice spacing has so far
been ignored, see e.g.~\cite{Hidaka:2012mz,Beane:2014ora,Detmold:2015daa}.

\section{QCD in an external magnetic field}

We consider an external magnetic field $\ve{B}=B \, \ve{e}_3$ pointing in
$x_3$-direction, which can be generated by a vector potential of the form
\begin{equation}
\label{eq:vector-pot}
A_1(x)=-\eta B x_2 \,, \quad A_2(x)=(1-\eta) B x_1 \quad \text{and} \quad
A_0(x)=A_3(x) = 0 \,.
\end{equation}
Here $\eta \in [0,1]$ is a free parameter which does not enter physical
observables. The following results are obtained using the symmetric gauge,
$\eta=1/2$, but we have explicitly checked that the results agree when we set
$\eta=1$. Note, that we also need to include ``twists'' at the boundary to satisfy
periodic boundary conditions for the gauge potential and that the flux of the
magnetic field is quantised according to $e B a^2 N_1 N_2 = 6\pi N_B$ ($0\leq
N_B<N_1 N_2$), where $N_i$ is the number of lattice points in $x_i$-direction
and $e>0$ is the elementary charge (see, e.g., Ref.~\cite{Bali:2011qj}). The
external magnetic field enters the system via the covariant derivative by
minimal coupling of $A_\mu$ to the quark charges $q_f$ as 
$D_\mu\to D_\mu+iq_fA_\mu$. For two flavours ($q_u=2e/3$ and $q_d=-e/3$), this
results in the replacement of the standard lattice link variables $U_\mu^{\rm
G}(x)\in\SU(3)$ in the Dirac operator by the new links (diagonal in flavour
space)
\begin{equation}
\label{eq:u3-links}
 U_\mu(x) = U_\mu^{\rm G}(x) u_\mu(x) \, \in \, \U(3)\times\SU_f(2) \quad
\text{with} \quad u_\mu(x)=\exp\Big[ i e a
\Big(\frac{1}{6}+\frac{\tau^3}{2}\Big) A_\mu(x)\Big]  .
\end{equation}
In the following we employ the unimproved Wilson Dirac operator $D_W$ with
hopping parameters  $\kappa_f$ corresponding to input quark masses
$a\bar{m}_f=(\kappa_f^{-1}-8)/2$.

\section{The free case}
\label{sec:free}

We begin the discussion with the free case, where quarks couple to the external
magnetic field through their electric charges but are blind to QCD interactions.
To find how the additive mass renormalisation for Wilson fermions depends on
$B$, we will make use of the particular form of the relativistic energy levels
(Landau-levels) for fermions ($|\ve s|=1/2$) and for bosons,
\begin{equation}
\textmd{fermions: }\;E^2_f(B) = m_f^2 + (2n+1)|q_fB| - 2\ve{s}\cdot q_f\ve{B}, \quad\;
\textmd{bosons: }\;E^2_f(B) = m_f^2 + (2n+1)|q_fB|.
\label{eq:LL}
\end{equation}

The massless Wilson Dirac operator may be written schematically as $aD_W =
a\slashed{D} + a^2\Delta/2$,  where $\slashed{D}$ is the anti-Hermitian part and
$\Delta$ the Wilson term. While the former describes spin-$1/2$ particles, the
latter is the discretisation of the Klein-Gordon operator and, thus, describes
bosons. This difference between spin-statistics becomes relevant when we switch
on the external magnetic field in both operators. Setting $m_f=0$ in
eq.~(\ref{eq:LL}), we read off that the lowest eigenvalue of $\slashed{D}$
vanishes ($n=0$ and $\ve{s} \cdot q_f\ve{B}>0$), while that of the Wilson term 
equals $a^2|q_fB|/2$. Accordingly\footnote{Even though $[\slashed{D},\Delta]\neq
0$ and thus the two operators do not share a common eigensystem, the lowest
eigenmodes do coincide, as can be checked analytically in the continuum. Thus,
the lowest eigenvalue of $aD_W$ is simply the sum of the lowest eigenvalues of
$a\slashed{D}$ and of $a^2\Delta/2$.}, the real part of the lowest eigenvalue 
of $D_W$ does not remain zero, but increases linearly with $B$. (We checked this
numerically by calculating the eigenvalues of $D_W$.) This increase is
equivalent to an additive shift in the quark mass, implying a $B$-dependent
additive renormalization. Thus, simulating at fixed quark mass $m_f$ is achieved
by tuning the hopping parameters along the trajectory,
\begin{equation}
\kappa_f^{-1}(B)=\kappa^{-1}(B=0) - c_f(B), \quad\quad\quad c_f(B) \equiv a^2|q_fB|.
\label{eq:free_tuning}
\end{equation}
Note that this additive shift in the spectrum is only present for Wilson
fermions and does not appear, for example, in the staggered
formulation~\cite{Endrodi:2014vza}.

To demonstrate the effect of this $B$-dependent tuning, it is instructive to 
measure the free ``pion masses'', i.e.\ the energies associated to the leading
decay of the pseudoscalar correlation functions, corresponding to the energies
of quark-antiquark states with imposed pion quantum numbers. On the one hand,
for neutral correlation functions,  both quarks have $\ve s \cdot q_f\ve B>0$ in
the ground state, so that the associated energy is $E(B)=m_f+m_{f'}$, cf.\
eq.~(\ref{eq:LL}). On the other hand, for the charged correlation function one
of the quarks (that with the smaller absolute charge) is forced to have $\ve s
\cdot q_f\ve B<0$, and the energy is given by $E(B)=m_f+\sqrt{m_{f'}^2+2|q_{f'}
B|}$. 

\begin{figure}[t]
\begin{minipage}[c]{.48\textwidth}
\centering
 \vspace*{-4mm}
\includegraphics[]{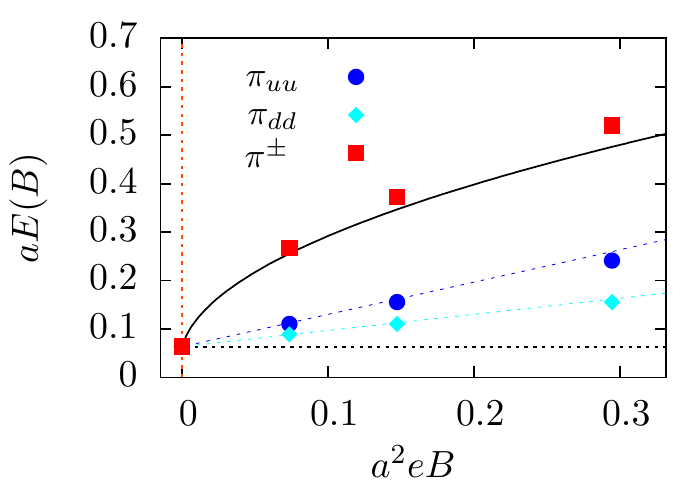}
\end{minipage}
\begin{minipage}[c]{.48\textwidth}
\centering
 \vspace*{-4mm}
\includegraphics[]{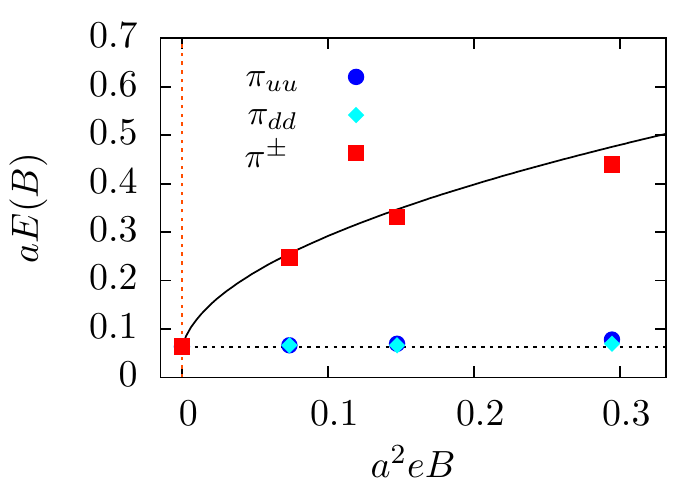}
\end{minipage}
 \vspace*{-1mm}
\caption{Results for the energies associated with free correlation functions in
the pseudoscalar channel without (left) and with (right) tuning $\kappa$
with $B$ along the trajectory~(\protect\ref{eq:free_tuning}). The coloured
dashed lines in the left plot show the analytic expectations
$E_f(B)=E(0)+c_f(B)$ for energies associated with neutral correlation
functions and the solid line is the expectation for charged ``pions''.}
\label{fig:free-pions}
\end{figure}

Our numerical results for the energies at fixed $\kappa=0.124$ are shown in
figure~\ref{fig:free-pions} (left). The energy of the neutral pion increases
with the magnetic field, indicating the unphysical increase of the quark mass by
the amount $c_f(B)$. Tuning the hopping parameters along the
trajectory~(\ref{eq:free_tuning}) instead, the neutral pion mass remains
constant as it should, see figure~\ref{fig:free-pions} (right). Our results are
also in good agreement with the expectation for the charged ``pion''.

\section{Full QCD}

\subsection{Additive quark mass renormalisation and Ward identities}

When the quarks feel the QCD interactions, the situation becomes more
complicated, since the neutral pion mass will change with the magnetic field.
However, it is known from chiral perturbation theory that even for non-zero
magnetic field the energy of the neutral pion remains zero at vanishing quark
mass~\cite{Agasian:2001ym}, so that a chiral extrapolation of these energies
allows to determine $\kappa_c$. In practice, the situation is even more involved
since the mass eigenstates are mixings between the flavour eigenstates and
$\kappa_c$ becomes flavour dependent, $\kappa_c\to\kappa_{c,f}(B)$. In addition,
the neutral correlation functions obtain a contribution from disconnected
diagrams, which we neglect.

An alternative is to use the current quark masses appearing in Ward-Takahashi
identities~\cite{Bochicchio:1985xa}. In the presence of external magnetic
fields the identities change, since the covariant derivatives are not
proportional to the unit matrix in flavour space. In the continuum the vector
and axial-vector Ward identities read (omitting the $U_A(1)$ anomaly term)
\begin{align}
 \label{eq:wi1}
 \partial_\mu (J_V)^j_\mu(x) = & i \epsilon_{3jk} \big\{ (m_u-m_d)
\bar{\psi}(x) \frac{\tau^k}{2} \psi(x) + i \bar{\psi}(x) \gamma_\mu A_\mu(x)
\frac{\tau^k}{2} \psi(x) \big\} \\
 \partial_\mu (J_A)^j_\mu(x) = & (m_u+m_d) \bar{\psi}(x) \gamma_5
\frac{\tau^j}{2} \psi(x) + \delta_{j3} \frac{1}{2} (m_u-m_d) \bar{\psi}(x)
\gamma_5 \ve{1} \psi(x) \nonumber \\ 
 \label{eq:wi2}
 & - \epsilon_{3jk} \bar{\psi}(x) A_\mu(x) \gamma_\mu
\gamma_5 \frac{\tau^k}{2} \psi(x) \,.
\end{align}
Here $J_{V/A}$ are vector and axial-vector currents, respectively, and
$\epsilon_{ijk}$ is the totally antisymmetric tensor.~\footnote{For the axial
Ward identity in the continuum and for domain wall fermions on the lattice
see~\cite{Blum:2007cy}.} On the lattice with Wilson fermions the Ward identities
obtain similar terms and additional dimension 5 operators due to the variation
of the Wilson term. A publication containing the lattice identities is in
preparation~\cite{future}.

For neutral correlation functions the new terms vanish, leaving the standard
identities in place. Naively, one can define singlett quark masses associated
with the connected neutral pions with $\bar{u}u$ and $\bar{d}d$ flavour content.
However, the ``neutral'' correlation functions in these identities receive
disconnected contributions. Furthermore, the axial identity for $j=3$ contains a
mixture of $\bar{u}u$ and $\bar{d}d$ correlation functions and their separation
demands the use of the identity for the unit matrix, which is violated at the
quantum level due to the axial anomaly. A clean alternative is to use the
identities for ``charged'' correlation functions. In this case no disconnected
diagrams contribute. The problem, however, is that vector and axial-vector
identities are needed to determine the point where $u$ and $d$ quark masses both
vanish. We plan to exploit this strategy in the future.

\subsection{A test of tuning with connected neutral correlation functions}
\label{sec:tuning}

Our quenched setup for testing uses a $48\times 16^3$ lattice at $\beta=6.00$.
Following~\cite{Bhattacharya:1995fz}, this results in a lattice spacing of
$a\approx0.09$~fm and the critical hopping parameter at $\ve{B}=\ve{0}$ is
$\kappa_c=1/(2a\bar{m}_c+8)=0.157131$. We have generated 200 uncorrelated
configurations and performed one measurement per configuration with a Wuppertal
smeared~\cite{Gusken:1989qx} source including spatially APE
smeared~\cite{Falcioni:1984ei} links, employing CHROMA~\cite{Edwards:2004sx}.

For a first determination of the shift in the additive quark mass
renormalisation in the interacting (quenched) case, we extract $\kappa_{c,f}(B)$
using the current quark masses $m^{\rm WI}_f$ associated with the Ward-Takahashi
identities including neutral correlation functions. We determine $m^{\rm WI}_f$
from
\begin{equation}
 \label{eq:PCAC-mass}
 m^{\rm WI}_f = \frac{\partial_0 \left< \big(J^{ff}_A\big)_0(x_0) P^{ff}(0)
\right>}{\left< P^{ff}(x_0) P^{ff}(0) \right>}
\end{equation}
for several values of $\kappa$ (we have results for neutral pion energies down
to 300 MeV for all values of $B$) and perform a linear chiral extrapolation to
the point where $m^{\rm WI}_f=0$ for each value of $B$. In
eq.~\refc{eq:PCAC-mass} $\partial_\mu$ is the backward lattice derivative, $J_A$
is the point-split axial vector current and the superscript $ff$ indicates
neutral operators. The details will be contained in our future publication.
Note, that we obtain fully compatible results when we extrapolate the connected
neutral pion energies instead.

\begin{figure}[t]
\begin{minipage}[c]{.48\textwidth}
\centering
 \vspace*{-4mm}
\includegraphics[]{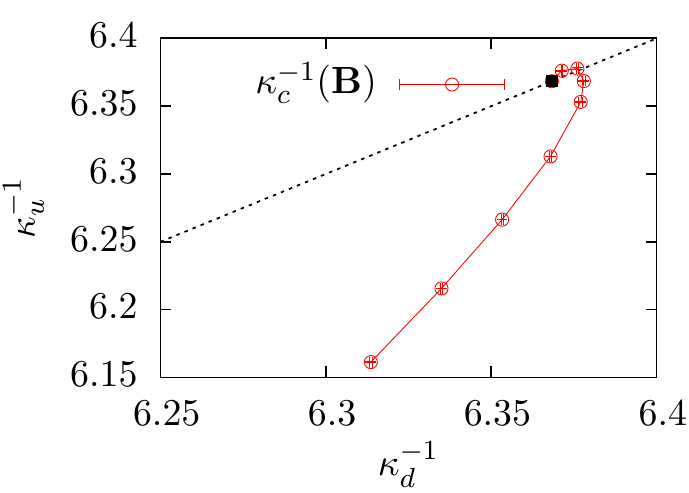}
\end{minipage}
\begin{minipage}[c]{.48\textwidth}
\centering
 \vspace*{-4mm}
\includegraphics[]{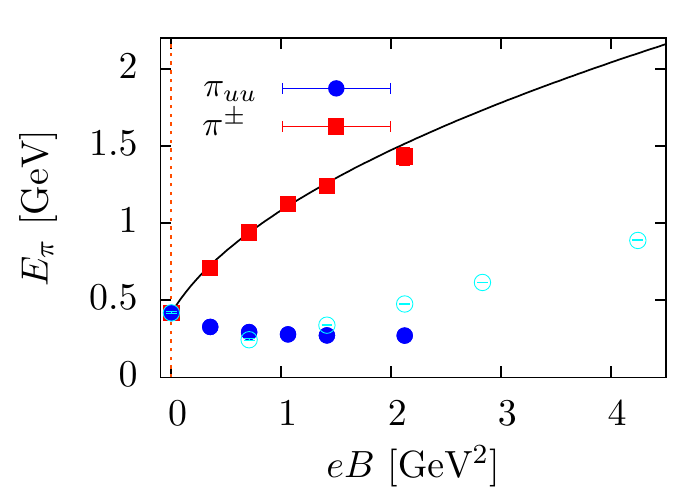}
\end{minipage}
 \vspace*{-1mm}
\caption{{\bf Left:} Results for the critical hopping parameters $\kappa_{c,f}$,
for $f=u,d$, for different values of $B$, as a trajectory in the plane of
inverse hopping parameters. The startpoint at $B=0$ is the black point on the
symmetric axis (dashed line). {\bf Right:} Results for the pion spectrum versus
the strength of the external magnetic field with $\kappa_u$ and $\kappa_d$ tuned
to achieve a constant bare quark mass. The black line shows the lowest
Landau level prediction for the charged pions and the open circles display the
$\pi_{uu}$ masses obtained from constant $\kappa$ for comparison.}
\label{fig:qmren}
\end{figure}

The results for $\kappa_{c,f}$ are shown as a trajectory in the plane of
inverse $u$ and $d$ quark hopping parameters in figure~\ref{fig:qmren} (left).
Their values reflect the non-monotonous behaviour of the neutral pion energies
seen in figure~\ref{fig:qmren} (right), leading to a curl in the trajectory.
Using these results we now retune the hopping parameter with $B$, to achieve a
constant bare quark mass, and look at the effect on the spectrum. The results
for the pion energies obtained with constant bare quark masses, corresponding to
a pion mass of about 400~MeV at $\ve{B}=\ve{0}$, are shown in
figure~\ref{fig:qmren} (right). For comparison we have also shown the energies
obtained when $\kappa=\kappa_u=\kappa_d$ is kept constant. Comparing the two
sets of results, we see that the retuning of the bare quark masses disposes of
the non-monotonous behaviour of the neutral pion masses and, consequentially,
the discrepancy with the results from other fermion discretisations.

\section{Conclusions and perspectives}

In this proceedings article we have reported on our ongoing study on the
spectrum of low lying hadrons exposed to strong external magnetic fields with
Wilson fermions. Figure~\ref{fig:pion-comp} shows a discrepancy in the large $B$
behaviour of connected neutral pions between results from Wilson fermions and
other fermion discretisations. We have shown that (at least parts of) this
discrepancy is due to a change in the additive renormalisation of the quark mass
$\bar{m}_{c,f}$ and the associated change in the bare quark mass with $B$. We
have shown that this effect is present already in the free case and that it
stems from the bosonic nature of the Wilson-term in the action. It can be
removed by tuning the hopping parameter as a function of $B$. The associated
change of the bare and renormalised quark mass affects all quantities that are
sensitive to changes in $m_f$ by (potentially large) lattice artefacts.
Accordingly, some caution is required when comparing quantities evaluated at
different magnetic fields at finite lattice spacing.

Since the pion energies change with the magnetic field, they cannot be used
for a direct tuning of quark masses. We thus propose to use the current
quark mass associated to lattice Ward-Takahashi identities in QCD+QED (the
Wilson fermion identities will be published in~\cite{future}) for the tuning and
reported on a first test using neutral correlation functions. Apart from changes
in the mutliplicative renormalisation, the associated current quark masses
provide a means to compare the renormalised quark masses at $B\neq0$ with the
ones at $B=0$. They can potentially also be used for a clean definition and
extraction of quark masses in QCD+QED.

\end{document}